\documentclass{PoS} 
\usepackage{amsmath}

\PoS{PoS(LAT2005)274} 
\title{Leibniz rule and exact supersymmetry on lattice: \\
 a case of supersymmetrical quantum mechanics} 

\ShortTitle{Leibniz rule  and exact supersymmetry  on lattice  }

\author{Mitsuhiro Kato\\ 

Institute of Physics, University of Tokyo, 
        Tokyo 153-8902, Japan\\ 
         E-mail: \email{kato@hep1.c.u-tokyo.ac.jp}} 
 
 \author{Makoto Sakamoto\\ 
 Department of Physics,
Kobe University, Rokkodai, Nada, Kobe 657-8501, Japan\\ 
         E-mail: \email{dragon@kobe-u.ac.jp}}

\author{\speaker{Hiroto So}\thanks{This work is supported in part by the 
Grants-in-Aid for Scientific Research No. 17043004-0167 from 
the Japan Society for the Promotion of Science.}\\
 Department of Physics,
        Niigata University, Ikarashi 2-8050, Niigata, 950-2181, Japan\\ 
         E-mail: \email{so@muse.sc.niigata-u.ac.jp}}

\abstract{Abstract\\
We propose a new formulation of lattice theory. 
It is given by a matrix form and suitable for satisfying  Leibniz rule 
on lattice. The theory may be interpreted  as a multi-flavor system. 
By realizing the difference operator as a commutator, we may obtain 
exact supersymmetric theories  on lattice explicitly. Some problems such as 
locality and single flavor reduction are also commented. }

\FullConference{XXIIIrd International Symposium on Lattice Field Theory\\ 
 
		 25-30 July 2005\\ 
 
		 Trinity College, Dublin, Ireland}

\begin{document}

\section{Motivations} 
 
Supersymmetry(SUSY) is necessary for understanding  unified field theories 
in  particle physics. The nonperturbative effects 
such as the vacuum property and SUSY breaking terms of unified theories 
take us back to our real world . 
To investigate the effects,  we must realize  full (or a part of) 
SUSY algebra in constructing exact SUSY on lattice.  
In the free theories, it is realized by a relation between fermion and 
boson propagators. 
On the other hand, a problem of Leibniz rule must be solved in 
the interacting theories\cite{D-N}. 

Our solution   for the problem 
 is a theory of matrix-valued fields and the realization of a difference operator 
 as a commutator. 
A  whole lattice is treated as a system just like a 'field'. 
After a decomposition of the matrix into a lattice field, 
it may be seen as a multi-flavor system. 

Kaplan {\it et al.} have approached  by imposing orbifolding conditions 
on a matrix  \cite{Kaplan1,Kaplan2,Kaplan3,Kaplan4}. But they did not define 
the difference operator in a meaning of matrix.  
In the context of noncommutative geometry, Bars and Minic presented 
a matrix-type theory\cite{Bars}.  They constructed a 2-dimensional theory 
on a matrix to obtain the non-commutative property. 

In this talk, we set a 1-dimensional lattice theory on a matrix, 
which covers more general construction ways. 
A Wess-Zumino model in a 1-dimensional case is constructed 
for simplicity. This is extendable to 2- and 4-dimensional 
cases. 

\section{Our Formalism} 

Our  formulation  is an extension of the ordinary lattice theory and 
to treat lattice fields as a whole.    
We consider a matrix-valued field    
$\phi_{ij}$,  where  $i,j=1, \cdots N $ and 
$N$ is a 1-dimensional lattice size. The product of matrix-valued fields 
follows to a usual matrix product rule.  
 The diagonal component, $\phi_{i,i}$ 
is recognized as a field on a site, $i$.  Furthermore, 
a field on a link is realized as  $\phi_{i,i+1}$. The trace operation 
corresponds to the space-time summation. 

The important key of our formulation is the definition of 
a difference operator. A commutator with a constant matrix, $d$ is identified 
as the  operator, 
\begin{equation}
i[d,\phi] \rightarrow \nabla \phi
\label{defofdiff}
\end{equation}
where $\nabla$ is a certain difference operator.

For the choice of $d$,  the following conditions are imposed;  

(1) hermitian, 

(2) local, 

(3) simple. 

\noindent
The explicit candidate for $d$ is 

\begin{equation}
d= \frac{1}{2i}(T^+ - T^-), 
\label{defofd}
\end{equation} 
\noindent
where $T^+$ is a forward displacement operator and 
 $T^-$ is a backward displacement operator; 

$$
T^+ =
\begin{pmatrix}
0 & 1 & 0 & \cdots & \cdots & 0\cr
 \vdots & \ddots & 1  & \ddots &\ddots & \vdots \cr
 \vdots & \ddots & \ddots & \ddots  & \ddots  & \vdots \cr
 \vdots & \ddots & \ddots & \ddots & 1 & 0 \cr
 0  & \ddots & \ddots & \ddots & \ddots & 1 \cr
 1  & 0 & \cdots & \cdots & \cdots & 0
\end{pmatrix}, 
~~~~~~~~~~~~~~~~T^- =
\begin{pmatrix}
 0 & \cdots & \cdots & \cdots  & 0 & 1 \cr
 1 & \ddots   & \ddots & \ddots  & \ddots & 0 \cr
 0 & 1   & \ddots& \ddots  & \ddots & \vdots\cr
 \vdots & \ddots  &\ddots & \ddots & \ddots&\vdots \cr
  \vdots  & \ddots & \ddots  & 1 & \ddots  & \vdots \cr
  0 & \hdots  & \hdots & 0 & 1 & 0  
\end{pmatrix}.
$$

Instead of taking the matrix form, our formulation may be 
interpreted as a multi-flavor system, which is useful in 
understanding  Leibniz rule on lattice.

A $(2N-1)$-flavored field is defined as

$$
 \phi_k(n) =  \phi_{i,j}, 
$$
\noindent
where a flavor index, $k\equiv j-i$ and a lattice site, 
$n\equiv i+j$. 
From Eq.(\ref{defofdiff}) and Eq.(\ref{defofd}), the difference operator  for a flavored field becomes to 

$$
\frac{1}{2}(\nabla^++\nabla^-)(\phi_{k-1} + \phi_{k+1}) ,
$$
\noindent
where $\nabla^+$ means a forward difference and $\nabla^-$ does a backward difference. 
It should be noted that the flavor indices are shifted. A matrix-valued theory and 
a multi-flavored system 
are summarized in Table 1.

\vspace{0.5 cm}

{\small Table 1.~ Correspondence among various theories. }
$$
\begin{array}{|c||c|c|c|}\hline
{\rm continuum ~theory} & {\rm lattice~ theory} & {\rm  {matrix-valued}~ theory}   
&{\rm  {multi-flavored}~ theory}   \\  \hline\hline
{\rm a~field} & \phi_n & \phi_{i,j} &  \phi_k(n) \\  \hline
{\rm coordinate} & n & i & n=i+j \\  \hline
{\rm flavor} & {\rm single~flavor} & {\rm single ~flavor} & (2N-1)-{\rm flavored~ system} \\ \hline
{\rm space ~integral} & \sum_n & {\rm tr} & \sum_{n}\\ \hline
{\rm a~ scalar ~ field}   & \phi_n & \phi_{i,i} &  \phi_{0} (n)  \\ \hline
\times   &  {\rm link ~ fields}~~ \phi_{n,\pm\hat{1}} & \phi_{i,i\pm1} &  \phi_{\pm1} (n)  \\ \hline
\times &  {\rm fields~ on~ double-length~ link} & \phi_{i,i\pm2} & \phi_{\pm2}(n) \\ \hline 
{\rm derivative} & {\rm difference} & i[d,\phi]& 
\frac{1}{2}(\nabla^++\nabla^-)(\phi_{k-1} + \phi_{k+1})  \\ \hline
\end{array}
$$

\section{Leibniz Rule} 

In this section, we clear the problem of Leibniz rule and  show how  
a commutator formulation or multi-flavor interpretation may solve the problem. 
A differential operator has generally the following properties: 
(1) linearity, (2) Leibniz rule, 
and (3) normalization or conjugation property for an independent variable. 
From the properties, we may differentiate any analytic function. 

The second  condition on lattice causes a problem.  
For example, 
$$
\nabla^+(\phi_n \psi_n) =  (\nabla^+ \phi_{n}) 
\psi_{n+1} +\phi_{n} (\nabla^+ \psi_n)  \ne  (\nabla^+ \phi_{n}) 
\psi_{n} +\phi_{n} (\nabla^+ \psi_n) .
$$
\noindent
The first identity is not suitable for constructing exact SYSY theories.  
This situation is caused under more general situation. 
The local lattice theory may not hold exact Leibniz rule. 
On the other hand, a commutation relation between matrices is always distributive. 
In  our formalism,  simple Leibniz rule holds,
$$
[d, \phi \psi] = [d, \phi] \psi +  \phi [d, \psi]  . 
$$
\noindent
In the multi-flavored system,  the shift of the flavor is essential. 
Some comments are put in order; for single flavor case,  Leibniz rule and  locality are 
    incompatible in lattice theory\cite{D-N}, 
our Leibniz rule for product of functions is similar to 
    that of non-commutative differential geometry\cite{Bars} .

\section{Interacting SUSY Theory} 

In the section, we construct an interacting SUSY model on lattice 
and represent exact SUSY. 
Although our explicit example is the model in 1 dimension,  higher dimensional cases are mentioned  in the summary.  We prepare a scalar matrix, $\phi$, 
fermi matrices, $\bar{\psi}, \psi$, and an auxiliary matrix, $F$. 

Our total action, $S$  is decomposed into the three parts 
on off-shell formalism;  

\begin{eqnarray}
S={\rm tr}~L= S_0+ S_1+ S_{{\rm int}},  
\label{action}
\end{eqnarray}
\noindent
where the free part is written as 
\begin{equation*}
S_0=
{\rm tr}~ (-\frac{1}{2}[d,\phi]^2 
-\frac{i}{2} (\bar{\psi}[d,\psi] 
- [d,\bar{\psi}]\psi) - \frac{1}{2} F^2) ,
\end{equation*}
\noindent
a doubling-avoiding (Wilson-like) action is  
$$
S_1= \alpha{\rm tr}~ (F\{M,\phi\} + \{\bar{\psi},M\} \psi),  
$$
\noindent
and  the interacting terms  $S_{\rm int}$ are 
\begin{eqnarray*}
S_{{\rm int}} = {\rm tr}~(\frac{\lambda}{2!}
F\phi^2 + \frac{\lambda}{2!}\sum_{s=0}^{1}\bar{\psi}
\phi^s\psi\phi^{1-s}).
\end{eqnarray*}
\noindent
It is noted that $\alpha$ is a nonzero constant and $M\equiv 2-T^+ -T^-$.

The action is invariant under  the following SUSY transformation with Grassman-odd
 parameters, $\epsilon$ and $\bar{\epsilon}$; 

\begin{equation}
\delta \phi = -(\bar{\psi} \epsilon +
\bar{\epsilon}\psi),
\label{SUSY1}
\end{equation}
\begin{equation}
\delta \psi = \epsilon(i[d,\phi]+F),
\label{SUSY2}
\end{equation}
\begin{equation}
\delta \bar{\psi} = \bar{\epsilon}(-i[d,\phi]+F) 
\label{SUSY3}
\end{equation}
\noindent
and 
\begin{equation}
\delta F = \epsilon (-i [d, \bar{\psi}])
 + \bar{\epsilon}(-i[d,\psi]).
 \label{SUSY4}
\end{equation}
\noindent
This transformation satisfies  an SUSY algebra;

$$
[\delta_{\bar{\epsilon}},\delta_{\epsilon}]
 {\cal O}
=2i\bar{\epsilon}\epsilon[d,{\cal O}] ,
$$
\noindent
where ${\cal O}$ is any field such as $\phi,~\psi,~\bar{\psi}$ and $F$.

The variation of the lagrangian, $L $  may be written as 

\begin{equation}
\delta L = [d, {\cal O}] ,
\label{varL}
\end{equation}
\noindent
where
$$
{\cal O}= \frac{1}{2}\bar{\psi}\epsilon[d,\phi]+
\frac{1}{2}\bar{\epsilon}[d,\phi]\psi+\frac{i}{2}(\bar{\epsilon}F\psi)
-\frac{i}{2}(\bar{\psi}\epsilon F)
$$
\begin{equation}
-i\alpha \bar{\epsilon}\psi\{M,\phi\}
-i\alpha \epsilon \bar{\psi}\{M,\phi\}-i\frac{\lambda}{2!}\epsilon\{\bar{\psi},\phi\}
\phi
-i\frac{\lambda}{2!}\bar{\epsilon}\{\psi,\phi\}\phi .
\label{Ope}
\end{equation}
\noindent
Owing to this ristricted writing space, we  have written these terms shorthandly. 
The actual expression  in Eqs. (\ref{action}) and (\ref{Ope}) 
must be careful in the  matrix-ordering.   
From Eq. (\ref{varL}), it is clear that our action is invariant under the transformation, 
\begin{eqnarray*}
\delta S = \delta {\rm tr~} L = {\rm tr}~[d,{\cal O}] = 0.
\end{eqnarray*}

\section{Summary} 

In this talk, we have presented a new formulation of lattice theory. 
This formulation is to consider a matrix as whole lattice field. 
Our proposal of matrix-valued field theory  is 
different from that of Kaplan {\it et al.}\cite{Kaplan1,Kaplan2,Kaplan3,Kaplan4} 
in points of   
the difference operator and the correspondence to the ordinary lattice fields. 
The operator  in our case is realized as a commutator with a constant matrix.    
The Leibniz rule  always holds in this formulation because the commutator has 
a distributative property. An explicit model with the exact SUSY has been 
constructed in 1 dimension.  
It is possible to extend to higher dimensional theories \cite{K-S-S} such as 
$$
\phi_{i,j} \rightarrow \phi_{(i_1,i_2),(j_1,j_2)}
$$
\noindent
in 2 dimensions. Non chiral theories  are realized in $D=2,4$ theories 
owing to avoiding doubling or Wilson-like terms \cite{K-S-S}.

The important problem is the continuum limit of our theory. 
Although the limit operation is a difficult problem, we must 
comment on a flavor-reduction and locality. 
Our  theory has a (2N-1)-flavored Wess-Zumino model 
in 1 dimension on lattice.  
The reduction to a single flavored model  must be done with keeping 
the exact SUSY invariance. 
The consistency between some constraints for matrices 
and  the invariance or the commutator with $d$ is an essential point. 
Nonlocal terms are generally induced from the reduction. 
Nevertheless, there are some circumstantial evidence for us to control its locality. 
The first evidence is  that the free part  of our action 
may be reduced to a single flavored system 
 with  only local terms.  The second is the transformation (\ref{SUSY1}), 
 (\ref{SUSY2}), (\ref{SUSY3}), (\ref{SUSY4}) 
 which is written by 
 only local terms. The final one is the existence of a local Nicolai map 
 in 1- and 2-dimensional models. 
The dynamical properties such as phase structure may find the expected reduction. 

In the remaining problems, 
inclusion of  gauge interactions for our theory and construction of chiral theories 
are  nontrivial matter specially.

 \end{document}